\documentclass[12pt,preprint]{aastex}

\newcommand{\sgr}{SGR~J1833$-$0832~}
\newcommand{\sgrnos}{SGR~J1833$-$0832}
\newcommand{\cxo}{{\it Chandra}~}

\begin{document} 
\vspace{0.8 in} 

\title{Discovery of a new Soft Gamma Repeater, \sgrnos} 

\author{
E.~G\"o\u{g}\"u\c{s}\altaffilmark{1},
G.~Cusumano\altaffilmark{2},        
A.~J.~Levan\altaffilmark{3},
C.~Kouveliotou\altaffilmark{4},
T.~Sakamoto\altaffilmark{5},
S.~D.~Barthelmy\altaffilmark{5},
S.~Campana\altaffilmark{6}, 
Y.~Kaneko\altaffilmark{1},
B.~W.~Stappers\altaffilmark{7},
A.~de~Ugarte~Postigo\altaffilmark{6},
T.~Strohmayer\altaffilmark{5},
D.~M.~Palmer\altaffilmark{8},
J.~Gelbord\altaffilmark{9},
D.~N.~Burrows\altaffilmark{9},
A.~J.~van~der~Horst\altaffilmark{4,10},
T.~Mu\~noz-Darias\altaffilmark{6}, 
N.~Gehrels\altaffilmark{5},
J.~W.~T.~Hessels\altaffilmark{10,11}, 
A.~P.~Kamble\altaffilmark{12},
S.~Wachter\altaffilmark{13},
K.~Wiersema\altaffilmark{14},
R.~A.~M.~J.~Wijers\altaffilmark{12}, 
P.~M.~Woods\altaffilmark{15}
}
 
\altaffiltext{1}{Sabanc\i~University, Orhanl\i-Tuzla, 34956 \.Istanbul, Turkey}
\altaffiltext{2}{IASF-Palermo/Istituto di Astrofisica Spaziale e Fisica Cosmica di Palermo/Istituto Nazionale di Astrofisica (INAF), 90146 Palermo, Italy }
\altaffiltext{3}{Department of Physics, University of Warwick, Coventry CV4 7AL, UK}
\altaffiltext{4}{Space Science Office, VP62, NASA/Marshall Space Flight Center, Huntsville, AL 35812, USA}
\altaffiltext{5}{NASA/Goddard Space Flight Center, Greenbelt, MD 20771, USA}
\altaffiltext{6}{INAF-Osservatorio Astronomico di Brera, Via E. Bianchi 46, I-23807 Merate (LC), Italy}
\altaffiltext{7}{Jodrell Bank Centre for Astrophysics, School of Physics and Astronomy, The University of Manchester, Manchester M13 9PL, UK }
\altaffiltext{8}{Los Alamos National Laboratory, B-244, P.O. Box 1663, Los Alamos, NM 87545, USA}
\altaffiltext{9}{Department of Astronomy \& Astrophysics, The Pennsylvania State University, 525 Davey Lab, University Park, PA 16802, USA }
\altaffiltext{10}{NASA Postdoctoral Program Fellow}
\altaffiltext{11}{Netherlands Institute for Radio Astronomy (ASTRON), Postbus 2, 7990 AA Dwingeloo, The Netherlands}
\altaffiltext{12}{Astronomical Institute, University of Amsterdam, Science Park 904, 1098 XH Amsterdam, The Netherlands}
\altaffiltext{13}{Spitzer Science Center/California Institute of Technology, Pasadena, CA 91125, USA}
\altaffiltext{14}{Department of Physics \& Astronomy, University of Leicester, Leicester, LE1 7RH, UK}
\altaffiltext{15}{Dynetics, Inc., 1000 Explorer Boulevard, Huntsville, AL 35806, USA}

\authoremail{ersing@sabanciuniv.edu} 

\begin{abstract} 

On 2010 March 19, the {\it Swift}/Burst Alert Telescope triggered on a short burst with temporal and spectral characteristics similar to those of Soft Gamma Repeater (SGR) bursts. The source location, however, did not coincide with any known SGR. Subsequent observations of the source error box with the {\it Swift}/X-ray Telescope and the Rossi X-ray Timing Explorer ({\it RXTE}) led to the discovery of a new X-ray source, with a spin period of 7.56 s, confirming \sgr as a new magnetar. Based on our detailed temporal and spectral analyses, we show that the new SGR is rapidly spinning down (4$\times$10$^{-12}$ s/s) and find an inferred dipole magnetic field of 1.8$\times$10$^{14}$ G. We also show that the X-ray flux of \sgr remained constant for approximately 20 days following the burst and then started to decline. We derived an accurate location of the source with the {\it Chandra} X-ray Observatory and we searched for a counterpart in deep optical and infrared observations of \sgrnos, and for radio pulsed emission with the Westerbork Radio Synthesis Telescope. Finally, we compare the spectral and temporal properties of the source to the other magnetar candidates. 

\end{abstract} 


\section{Introduction} 

The first confirmed Soft Gamma Repeater (SGR) burst was detected in 1979 but it took $\sim7$ years (\citealt{laros87,atteia87,kouv87}) to associate it with a newly discovered population of astronomical objects, namely neutron stars repeatedly emitting soft and short gamma-ray bursts with peculiar properties. A decade later, these objects, together with their cousins, Anomalous X-ray Pulsars (AXPs; see e.g., \citealt{mer08}), were shown to possess extremely high magnetic fields ($\sim10^{15}$ G), as deduced from their spin periods and spindown rates, and were named magnetars \citep{dt92, td96, kouv98, vas97}.  The most distinguishing magnetar characteristics are a narrow range of spin periods (2$-$12s), high spindown rates ($\sim10^{-10}$ s/s) and unpredictable burst-active episodes. The latter can have a multitude of bursts with a wide luminosity range, anywhere from 10$^{-2}$ to 10$^7$ times the Eddington luminosity. It is these bursts that enable the initial classification of a source as an SGR, to be subsequently confirmed as a magnetar with timing analysis (see \citealt{woo06} and \citealt{mer08} for a detailed review on magnetars.)

By 2008 a dozen sources were confirmed magnetars and a couple more were listed as magnetar candidates pending observations of new burst active episodes. For the known sources, frequent spin period monitoring observations with the {\it Rossi} X-ray Timing Explorer ({\it RXTE}) over the last fifteen years have provided a wealth of insight about their dynamics and energetics. Interestingly, despite their common characteristics, there are several important differences across this small population, and a larger sample of objects at different evolutionary stages is necessary to better understand the phenomenon. Magnetar hunting however, is not an easy task; we have two strikes against us {\it ab initio}:  their very small numbers and their very short active periods (days to months). Thus, the best way to discover magnetars is the combination of an all-sky X-and gamma-ray monitor -- to detect their bursts -- and a timing instrument -- to measure their spin and spindown rates, and estimate their magnetic fields. 

On 2010 March 19, the Burst Alert Telescope (BAT) onboard NASA's  {\it Swift} satellite triggered at 18:34:50 UT on one short and soft burst (Barthelmy et al. 2010) originating from a region in the plane of our Milky Way galaxy that harbors at least three other magnetar candidates: AXP XTE J1810-197, and SGRs 1806-20 and 1801-23. Using the initial source localization with the BAT, which had a  3$\arcmin$ error radius, we were able to excluded all these sources; further, based on the spectral and temporal properties of the event, we identified the source of the emission as a possible new SGR. Subsequent observations (starting from 60~s after the BAT trigger) of the source region with the {\it Swift}/X-Ray Telescope (XRT) provided the source position with a 1.6$\arcsec$ uncertainty \citep{gel10}. {\it Swift}/Ultraviolet Optical Telescope (UVOT) observations starting 71 s after the trigger did not detect a source in the UV or optical bandpass at the XRT position \citep{margel10}. To confirm the nature of the source, we observed it with {\it RXTE} for 1.3 ks {\it less than 3.25 hours} after its discovery was reported. A subsequent timing analysis of this observation revealed coherent pulsations in the X-ray flux of the source with a period of 7.56 s \citep{gogus10a}. Thus, we were able to identify \sgr as a slowly rotating neutron star in less than 6 hours after its discovery, the second time such a fast discovery was enabled only due to the unprecedented synergy of {\it Swift} and {\it RXTE} (the first was the identification of SGR J0501+4516; \citealt{gogus09}). 

{\it Swift} was the only satellite that triggered on \sgr; although the source was in the field of view of the Gamma ray Burst Monitor (GBM) onboard the Fermi Gamma-ray Space Telescope ({\it FGST}), the fluence of the event was much lower than any of the GBM trigger thresholds. To precisely locate the source we triggered our Target of Opportunity (ToO) observations with the {\it Chandra} X-Ray Observatory ({\it CXO}) and obtained observations with the European Southern Observatory ({\it ESO}) Very Large Telescope/High Accuity Wide-field K-band Imager ({\it VLT/HAWKI}) and with the Gran Telescopio Canarias/Optical Imager and Spectrograph ({\it GTC/OSIRIS}). We also observed with the Westerbork Synthesis Radio Telescope Pulsar Machine II ({\it WSRT/PuMa II}) and searched for radio pulsations from the source. 

In this paper, we report on the discovery of \sgrnos, and introduce its characteristics through our detailed multi-wavelenth investigations. In \S 2, we describe all multi-wavelength data used in our study. We present in \S 3 the precise localization of the new source using {\it CXO}, 2MASS and UKIRT data, and the temporal and spectral characteristics of the persistent X-ray emission from the source using our {\it Swift} and {\it RXTE} observations. Finally, we discuss and interpret our results in \S 4.

\section{Observations}

\subsection{Swift/BAT}

Following the 2010 March 19 BAT discovery of a burst (trigger=416485; \citet{gel10,bar10}, {\it Swift} slewed immediately to the on-board BAT position of RA, Dec = 18$^{\rm h}$33$^{\rm m}$46$^{\rm s}$, -08$^\circ$32$\arcmin$13$\arcsec$ ({\it l, b} = 23$^{\rm h}$19$^{\rm m}$38$^{\rm s}$, 0$^\circ$ 0$\arcmin$ 22$\arcsec$; radius uncertainty of $3\arcmin$, 90\% containment). This location is 13.3$^\circ$ from the position of SGR~1806$-$20. The {\it Swift}/XRT detected a new X-ray source within the BAT error circle; these observations are described in \S\ref{sec:swift_xrt}. This was the only triggered burst detected from the source with BAT and we report below on its spectral and temporal characteristics.  

\begin{figure}[ht]
\centerline{
\includegraphics[angle=-90,width=6.0in]{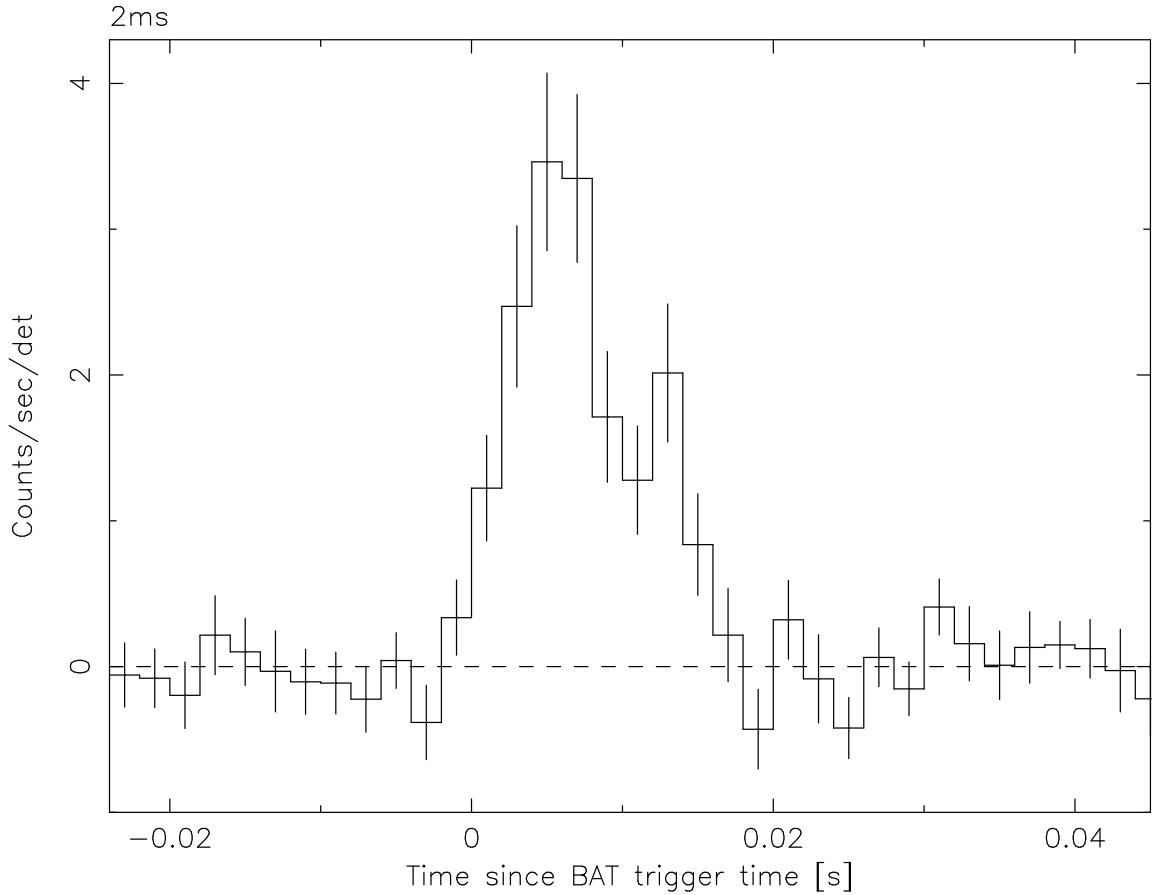}}
\caption{The {\it Swift}/BAT lightcurve of the burst from \sgr in the $15-350$ keV band. The time bin size is 2 ms.}
\label{fig:bat_lc}
\end{figure}

The burst was detected within the 87\% BAT partial coding region; its mask-weighted light curve (Figure \ref{fig:bat_lc}) shows a single peak structure. The $T_{\rm 90}$ duration \citep{kouv93} of the burst is 0.013 $\pm$ 0.002 s ($15-350$ keV; total error including systematics). 

We found that several continuum models could provide a statistically acceptable fit to the time-integrated spectrum ($T_0$ to $T_0+$0.016 s); a power law with an exponential cutoff \footnotemark
\footnotetext{Power law with an exponential cutoff is parametrized as
$dN/dE = A (E/50\,{\rm keV})^{-\alpha} \exp{(-E(2-\alpha)/E_{\rm peak})}$} with a photon index, $\alpha$, of $-1.02 \pm 1.35$ and an $E_{\rm peak}$ of $38.2 \pm 5.0$ keV ($\chi^2$/degrees of freedom (d.o.f.) = 43.5/56), a blackbody with $kT_{BB}= 10.0 \pm 1.2$ keV ($\chi^2$/d.o.f. = 43.8/57), and an optically thin thermal bremsstrahlung\footnotemark \footnotetext{OTTB is parametrized as $dN/dE = A (E/50\,{\rm keV})^{-1} \exp{(-E/kT)}$} with $kT_{\rm OTTB}= 29 \pm 5$ keV ($\chi^2$/d.o.f = 52.2/57). For the cutoff power law model, the total fluence in the $15-150$ keV band is $(1.3 \pm 0.2)\times 10^{-8}$ erg/cm$^2$ and the 6 ms peak flux (measured from $T_0+2$ to $T_0+8$ ms) is (9.9$\pm 2.1$)$\times 10^{-7}$ erg/cm$^2$/s ($19.4\pm 3.8$ photons/cm$^2$/s). All quoted errors are at the 90\% confidence level.

\subsection{Swift/XRT}\label{sec:swift_xrt} 

{\it Swift}/XRT observations of \sgr were obtained as a ToO monitoring program starting on 2010 March 19 at $\sim10$ ks per observation for a total exposure of $\sim$184 ks (Table~\ref{sgr:tab:xrtobs}). The XRT observations were coordinated with {\it RXTE} in a manner that complemented and optimized usage of both instruments. The XRT data were accumulated in Windowed Timing (WT) mode only during the third observation (002, see Table~\ref{sgr:tab:xrtobs}), while all other data were collected in Photon Counting (PC) mode. In the WT mode only the central $8\arcmin$ of the field of view is read out, providing one-dimensional imaging and full spectral capability with a time resolution of 1.8 ms. The PC mode provides full spatial and spectral resolution with a time resolution of 2.5 s.

The data were processed with standard procedures ({\sc xrtpipeline} v0.12.1) and filtering and screening criteria, using {\sc FTOOLS} in the {\sc Heasoft} package (v.6.6.1). We used a 0--2 and 0--12 grade selection for data in the WT and PC modes, respectively. This selection provides the best combination of spectral resolution and detection efficiency. \sgr was imaged far from the CCD hot columns so that no corrections inside the photon extraction regions were necessary.

We extracted the source events for timing and spectral analysis from a circular region of $47\arcsec$ radius (20 pixels) centered on the source position as determined with {\sc xrtcentroid}. 
The background was extracted from an annular region centered on the source with $90\arcsec$ and $120\arcsec$ internal and external radii, respectively. Ancillary response files were generated with {\sc xrtmkarf}, and account for different extraction regions, vignetting, and Point Spread Function corrections. We used the spectral redistribution matrices v011 in CALDB. All spectral fits were performed using {\sc xspec} v.12.5.10 (Arnaud 1996).

\begin{table}
\begin{center}
\caption{Summary of the {\it Swift}/XRT observations.\label{sgr:tab:xrtobs}}
\begin{tabular}{llll}
\hline
\noalign{\smallskip}
Seq.     & Start time  (UT)  & End time   (UT) & Exposure    \\ 
         &                   &                 &  (s)        \\
  \noalign{\smallskip}
 \hline
\noalign{\smallskip}
000     &  2010-03-19 18:35:50     &       2010-03-20 15:38:07     &       29174    \\
001     &  2010-03-21 04:14:14     &       2010-03-21 15:41:31     &       10756    \\
002     &  2010-03-21 16:47:37     &       2010-03-22 02:32:06     &       9954     \\
003     &  2010-03-22 07:50:12     &       2010-03-22 20:48:15     &       9869     \\
004     &  2010-03-23 08:58:47     &       2010-03-23 23:39:46     &       8068     \\
005     &  2010-03-24 01:07:40     &       2010-03-25 00:00:13     &       10266    \\
006     &  2010-03-25 01:36:34     &       2010-03-26 00:00:12     &       9844     \\
007     &  2010-03-26 12:28:42     &       2010-03-26 22:33:11     &       9961     \\
008     &  2010-03-27 10:57:33     &       2010-03-27 21:02:31     &       9855     \\
009     &  2010-03-28 00:04:08     &       2010-03-28 18:02:39     &       10919    \\
010     &  2010-03-29 12:44:13     &       2010-03-29 21:13:35     &       9511     \\
011     &  2010-03-31 00:20:50     &       2010-03-31 23:11:58     &       7942     \\
012     &  2010-04-03 14:59:40     &       2010-04-03 23:29:26     &       10028    \\
013     &  2010-04-07 13:45:23     &       2010-04-07 22:15:53     &       10097    \\
014     &  2010-04-12 09:41:24     &       2010-04-12 22:42:57     &       5130    \\
015     &  2010-04-13 03:08:11     &       2010-04-13 08:21:51     &       4054    \\
016     &  2010-04-15 00:01:07     &       2010-04-15 11:23:58     &       9417    \\
017     &  2010-04-18 14:53:02     &       2010-04-19 07:10:56     &       8817    \\

\noalign{\smallskip} 
\hline
\end{tabular}
\end{center}
\end{table}

\subsection{{\it RXTE}}

We monitored the source with the {\it RXTE}/Proportional Counter Array (PCA) starting on 2010 March 19, 21:49:10 UT (i.e., $\sim$3.25 hrs after the BAT trigger) until 2010 May 6 in 31 pointings. The time spacing between consecutive observations was on average 1.6 days, ranging between 0.06 and 4.9 days. The exposure of individual observations varied between 1.3 ks and 12 ks, with a total exposure of 178.6 ks over 47 days. In Table \ref{sgr:tab:rxteobs}, we list a log of RXTE observations of \sgrnos.

In this study we employed PCA data collected in the $2-10$ keV range. We inspected the lightcurve of each observation binned with 0.03125 ms to search for additional SGR bursts, and identified four events with duration $\lesssim$12 ms. These events were rather weak compared to the typical bursts seen with the PCA from other SGRs; their detection indicates that a low level of bursting activity continued after the BAT trigger. We then filtered out the times of these four short bursts to obtain burst-free event lists for our timing studies. Finally, we converted all event arrival times to the Solar System Barycenter.

\begin{table}
\begin{center}
\caption{Summary of the {\it RXTE} observations.\label{sgr:tab:rxteobs}}
\begin{tabular}{llll}
\hline
\hline
\noalign{\smallskip}
ObsID           & Start Time (UT)        & End Time (UT)          & Exposure (s) \\
\noalign{\smallskip}
\hline
\noalign{\smallskip}
95048-03-01-10  &  2010 Mar 19 21:49:10  &  2010 Mar 19 22:19:10  &  1335	 \\
95048-03-01-00  &  2010 Mar 21 08:07:28  &  2010 Mar 21 12:13:14  &  10443       \\
95048-03-01-01  &  2010 Mar 21 12:50:24  &  2010 Mar 21 15:22:10  &  6892	 \\
95048-03-01-02  &  2010 Mar 21 17:32:32  &  2010 Mar 21 23:13:10  &  13396       \\
95048-03-01-04  &  2010 Mar 23 05:56:32  &  2010 Mar 23 11:03:10  &  11936       \\
95048-03-01-03  &  2010 Mar 23 15:03:28  &  2010 Mar 23 19:08:10  &  9719	 \\
95048-03-01-05  &  2010 Mar 24 08:19:28  &  2010 Mar 24 09:17:10  &  3425	 \\
95048-03-01-07  &  2010 Mar 24 11:27:28  &  2010 Mar 24 12:19:10  &  3071	 \\
95048-03-01-08  &  2010 Mar 25 07:52:32  &  2010 Mar 25 10:21:10  &  6723	 \\
95048-03-02-00  &  2010 Mar 26 04:25:20  &  2010 Mar 26 06:49:10  &  6451	 \\
95048-03-02-01  &  2010 Mar 27 03:55:28  &  2010 Mar 27 06:21:10  &  6562	 \\
95048-03-02-02  &  2010 Mar 28 15:59:28  &  2010 Mar 28 18:18:08  &  5412	 \\
95048-03-02-03  &  2010 Mar 29 23:19:28  &  2010 Mar 30 01:39:10  &  5960	 \\
95048-03-02-04  &  2010 Mar 30 22:51:28  &  2010 Mar 31 01:02:10  &  5616	 \\
95048-03-02-05  &  2010 Apr  1 06:27:28  &  2010 Apr  1 07:12:10  &  2633	 \\
95048-03-02-06  &  2010 Apr  1 07:50:24  &  2010 Apr  1 08:45:10  &  3283	 \\
95048-03-03-00  &  2010 Apr  2 02:39:28  &  2010 Apr  2 05:10:08  &  6804	 \\
95048-03-03-01  &  2010 Apr  3 00:37:20  &  2010 Apr  3 01:34:10  &  3341	 \\
95048-03-03-06  &  2010 Apr  3 02:11:28  &  2010 Apr  3 03:08:10  &  3371	 \\
95048-03-03-02  &  2010 Apr  3 21:01:20  &  2010 Apr  3 23:32:10  &  6774	 \\
95048-03-03-05  &  2010 Apr  6 19:38:24  &  2010 Apr  6 20:35:10  &  3321	 \\
95048-03-04-00  &  2010 Apr  9 02:33:20  &  2010 Apr  9 04:34:48  &  6810	 \\
95048-03-04-01  &  2010 Apr 13 00:58:24  &  2010 Apr 13 03:01:10  &  5116	 \\
95048-03-05-00  &  2010 Apr 18 21:55:28  &  2010 Apr 19 00:26:10  &  6824	 \\
95048-03-06-00  &  2010 Apr 23 18:03:28  &  2010 Apr 23 20:35:10  &  6949	 \\
95048-03-06-01  &  2010 Apr 26 15:05:20  &  2010 Apr 26 15:53:10  &  2831	 \\
95048-03-06-02  &  2010 Apr 26 16:40:32  &  2010 Apr 26 17:36:10  &  3309	 \\
95048-03-07-00  &  2010 Apr 30 13:14:24  &  2010 Apr 30 15:45:10  &  6892	 \\
95048-03-07-03  &  2010 May  3 11:50:24  &  2010 May  3 12:46:08  &  3279	 \\
95048-03-07-01  &  2010 May  3 13:24:32  &  2010 May  3 14:21:10  &  3347	 \\
95048-03-07-02  &  2010 May  6 10:27:28  &  2010 May  6 12:58:10  &  6773	 \\
\noalign{\smallskip}
\hline
\end{tabular}
\end{center}
\end{table}

\subsection{{\it Fermi}/GBM}

Although GBM did not trigger on the {\it Swift}/BAT burst, we searched the daily GBM data (with time resolution of 256 ms in continuous mode and 64 ms in trigger mode) using our untriggered event search algorithm \citep{kan10} to uncover any additional weak events from the source. The search was performed in the 10$-$300 keV band on a 20-day long segment starting from 2010 March 15 (i.e., 4 days prior to the BAT trigger). We did not find any untriggered events originating from the direction of \sgrnos, including the BAT-triggered burst and the bursts detected with RXTE. This is most likely due to the extremely short duration of the bursts ($\lesssim16$\,ms), which made them practically undetectable over the background with the coarse time resolution of the GBM continuous data. We also inspected the GBM data for the times of the four short and weak {\it RXTE} bursts and again found no evidence of activity.

\subsection{{\it Chandra}}\label{sec:chandraobs}

We initiated ToO observations with the {\it Chandra} Advanced CCD Imaging Spectrometer (ACIS-I) in the Timed Exposure (TE) mode on 2010 March 23 for an effective exposure of 33 ks. The source was relatively bright at the time of the ACIS-I observation, resulting in a (background-subtracted) count rate of 0.104 $\pm$ 0.003 counts/s ($0.5-8$ keV), which is expected to cause a pile-up at the $\sim$13$\%$ level. Our instrument choice was based on our goal, which was primarily to determine an accurate X-ray position of the source, as described in \S\ref{sec:position}. We selected data collected in the 0.5$-$8.0 keV band with all four ACIS-I chips and generated a binned image (by a factor of 2) to improve the signal to noise ratio. We then employed wavdetect\footnotemark \footnotetext{A source finding algorithm in CIAO 4.0} to search for sources with a detection threshold of 5$\sigma$ above background and identified six uncatalogued point sources, including \sgrnos. We then repeated the same source finding procedure with an unbinned image to determine the accurate X-ray position of the SGR, which was at the aim point of ACIS-I. We list the X-ray positions (J2000) of all sources in Table \ref{tbl:acisi}. The last column in the table indicates whether an IR counterpart was identified for a source in the Two Micron All Sky Survey (2MASS) data (see also \S\ref{sec:position}).

\begin{table}[ht]
\begin{center}
\caption{X-ray point sources in the \cxo ACIS$-$I field of \sgrnos.\label{tbl:acisi}}
\begin{tabular}{ccccc}
\hline
Source  &  RA (J2000)   & Dec (J2000)  & S/N  & 2MASS \\
Name    & (hh mm ss.ss) & (dd mm ss.ss)& Ratio & Association \\
\hline
CXOU J183344.4-083108  &  18 33 44.38  & -08 31 07.71  &  53.7   &  SGR, No\\
CXOU J183320.2-083426  &  18 33 20.17  & -08 34 26.34  &   6.1   &  Yes  \\
CXOU J183322.7-082409  &  18 33 22.72  & -08 24 08.89  &   6.5   &  Yes  \\
CXOU J183317.9-082516  &  18 33 17.86  & -08 25 15.45  &   6.1   &  Yes  \\
CXOU J183316.2-083615  &  18 33 16.22  & -08 36 15.22  &   8.7   &  Yes  \\
CXOU J183359.4-082229  &  18 33 59.44  & -08 22 29.18  &   6.4   &  No   \\
\hline
\end{tabular}
\end{center}
\end{table}

\subsection{Optical and IR Observations}

The field of \sgr was observed with the imager OSIRIS, mounted on the 10.4m Gran Telescopio Canarias (GTC) telescope, starting on 2010 March 20 at 05:41:13 UT (i.e., 11.11 hrs after the BAT trigger). The observations were performed under non-optimal atmospheric conditions, with an average seeing of $\sim1.6\arcsec$, and consisted of 20$\times$90 s long exposures of a $1.85\arcmin \times 2.35\arcmin$ portion of the sky using a Sloan z-band filter. The left image of Figure \ref{fig:image_gtc_vlt} shows the combined GTC image of all 20 exposures of the field around \sgrnos. The mean time of the observations is 06:04:04 UT on 2010 March 20. We do not detect any optical source within the \cxo error circle in the combined image down to a 3$\sigma$ limiting magnitude of 24.9. 

We also observed the source using the {\it ESO VLT}/HAWKI in the Ks-band under excellent seeing conditions (0.45\arcsec). A total of 44 dithered exposures of duration 60 seconds were obtained, providing a total exposure time of 2640 s. The mean time of the observations was 09:01:09 UT on 2010 March 25. There is no obvious IR source within the \cxo error circle; further astrometry and limiting magnitude information is given below (\S\ref{sec:position}).

\subsection{{\it WSRT/PuMa~II} Radio Observations}

We observed \sgr on three epochs in March, all within a week of the source discovery, using {\it WSRT} at central frequencies of 2280 and 1380\,MHz (see Table \ref{tab:wsrtobs} for details). The pointing position was $RA = 18^{\rm h}33^{\rm m}44^{\rm s}.4, Dec = -08^\circ31\arcmin08\arcsec.4$. {\it WSRT} was used in its tied-array mode, but as the position of the source was relatively uncertain, only telescopes out to baselines of 1.5\,km were used. Furthermore, since one telescope was unavailable, only 11 telescopes were combined, giving a resolution of approximately $15\arcsec$ and $24\arcsec$ for 2280 and 1380 MHz, respectively, and a gain of $\sim$1\,K/Jy. The {\it WSRT} splits the 160 MHz bandwidth into 8 bands of 20 MHz wide, and each band is recorded at baseband using the {\it PuMa~II} pulsar backend \citep{karuppusamy2008}. The baseband data were converted into filterbank format with 64 channels per 20 MHz band and a sampling time bin of 8.192 ms. The resulting 512-channel filterbank data were searched blindly for a periodicity matching the known period of \sgrnos, and also folded using an ephemeris derived from the X-ray data (\S\ref{sec:timingrxte}). We searched over a range of dispersion measures of 0--1800\,pc\,cm$^{-3}$, where the maximum is that expected along the line of sight through the Galaxy in the NE~2001 model \citep{cordes2002}. The long spin period of the source means that special attention had to be given to identify potential radio frequency interference. However, while there were some long period features present in the data, they did not significantly affect our sensitivity to the source. Inspection of both the folded pulse profiles and the candidate lists from the blind searches did not reveal any significant detections of \sgrnos. Using the radiometer equation \citep{dewey1985} and considering pulse widths of 50\% and 10\%, which span the range of likely pulse widths for radio emitting magnetars, we obtain sensitivity limits of 0.6 and 0.2 mJy at 2280 MHz and 0.3 and 0.1 mJy at 1380 MHz, corresponding to the 50\% and 10\% widths, respectively.

\begin{table}[h]
\caption{{\it WSRT $-$ PuMa II} Observation Log}\label{tab:wsrtobs}
\begin{center}
\begin{tabular}{cccc} 
\hline
Epoch  &  MJD (day)  &  Duration (hour)  &  Frequency (MHz) \\
\hline
23 March 2010  & 55278.1570 &  1  &  2280  \\
23 March 2010  & 55278.1985 &  1  &  1380  \\
24 March 2010  & 55279.1962 &  1  &  1380  \\
26 March 2010  & 55281.2034 &  1  &  1380  \\
\hline
\end{tabular}
\end{center}
\end{table}

\section{Data Analyses and Results} 

\subsection{Precise Source Location and IR Counterpart upper limits}\label{sec:position}
 
\subsubsection{Imaging and astrometry}
 
\begin{figure}[ht]
\centerline{
\includegraphics[angle=-90,width=6.0in]{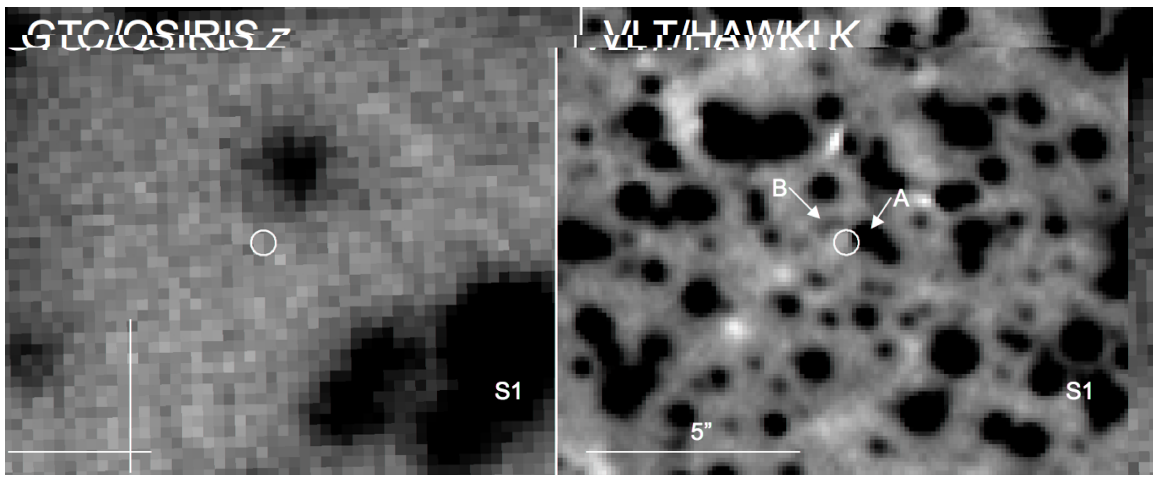}}
\caption{A zoomed in region around \sgr in the  {\it GTC/OSIRIS} image (left) and in the {\it VLT/HAWKI} image (right), showing the \cxo position (1$\sigma$ error radius). Labels {\it A} and {\it B} indicate  the objects close to the \cxo position of \sgr (see \S 3.1.2).}
\label{fig:image_gtc_vlt}
\end{figure}

\begin{figure}[ht]
\centerline{
\includegraphics[angle=0,width=6.0in]{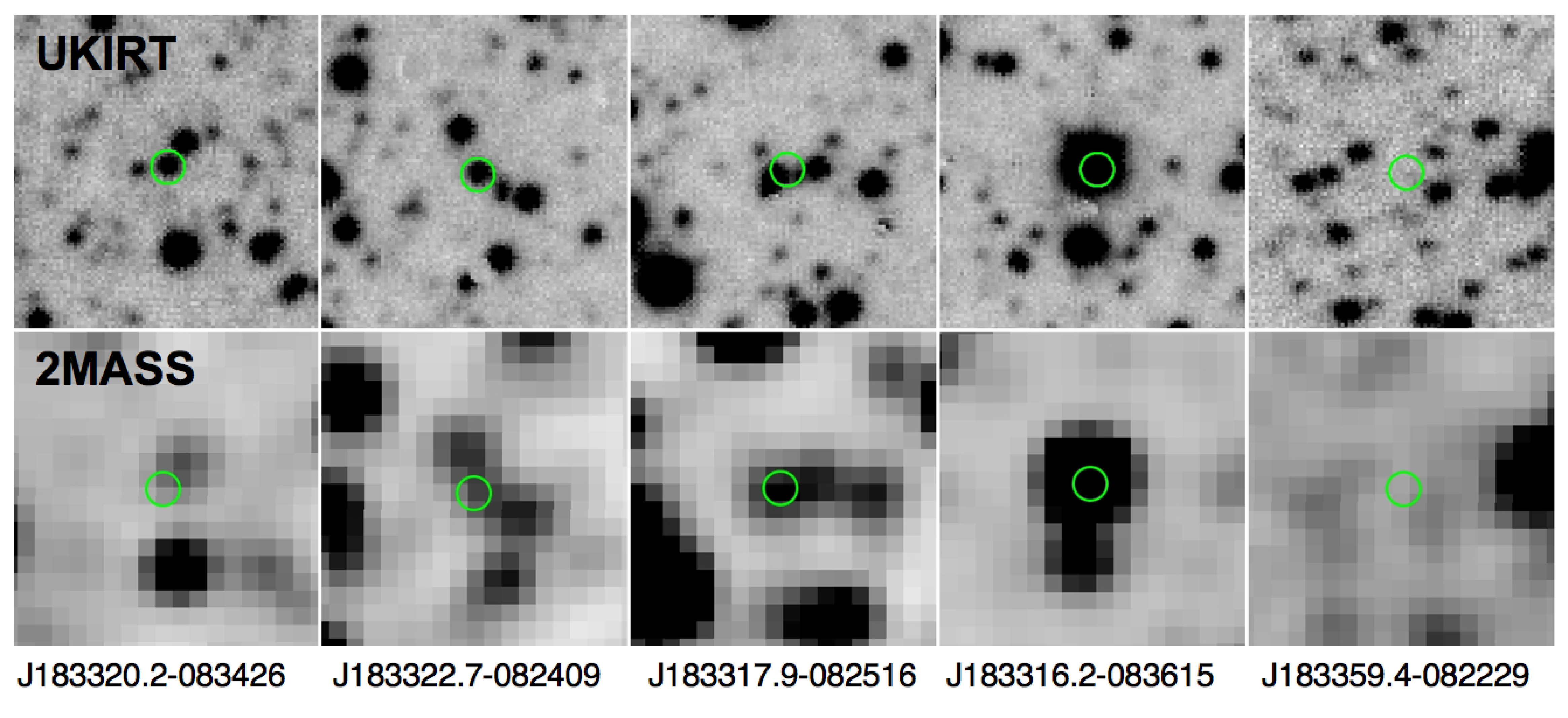}}
\caption{\cxo locations for five X-ray sources overlayed on {\it UKIRT} images (top) and 2MASS images (buttom). As can be seen the {\it UKIRT} allows an additional two counterparts to be identified (CXOU J183320.2$-$083426 and CXOU J183322.7$-$082409), which were ambiguous in the 2MASS data.}
\label{fig:sgr_cxo_ukirt}
\end{figure}

We searched the 2MASS archival data for IR counterparts of the five X-ray sources significantly detected in the {\it Chandra} image (\S\ref{sec:chandraobs}), in addition to \sgr itself. To perform astrometry, we utilize observations taken as part of the United Kingdom Infra-Red Telescope ({\it UKIRT}) Deep Sky Survey ({\it UKIDSS}), which obtains deeper images than 2MASS (limiting magnitudes of K$\sim 18$) with an improved PSF to aid with the elimination of source confusion. We show this improvement graphically in Figure \ref{fig:sgr_cxo_ukirt}, where we plot 2MASS and UKIRT images of each of the Chandra X-ray sources (excluding SGR 1833-0832, which is shown in Figure \ref{fig:image_gtc_vlt}).  We clearly see that one source, confused in the 2MASS images (CXOU J183320.2-083426), can be resolved by UKIRT; a second source (CXOU J183317.9-082516) remains apparently confused in the UKIRT images due to the contribution of a bright and faint source within the Chandra position. For the two remaining sources (CXOU J183320.2-083426 and CXOU J183316.2-083615) we confirm associations with 2MASS sources. To obtain astrometry of the field we first aligned the {\it UKIRT} data to 2MASS, which gives a fit with an RMS of $\sim 0.20\arcsec$.
Because of the source confusion in the 2MASS images at faint magnitudes, we limit ourselves to stars in the magnitude range $8 < K < 12$. We then compared the predicted locations of the X-ray sources with their measured centroids, and fit the resulting data with a simple linear shift in $x$ and $y$ coordinates (since we do not have a large number of sources for comparison this method provides a better estimate than allowing a larger number of free parameters). The resulting shifts are small: -0.25$\arcsec$ in RA and 0.2$\arcsec$ in Dec. The position of \sgrnos, accurate to 0.40$\arcsec$ is, $RA = 18^{\rm h}33^{\rm m}44^{\rm s}.37$, $Dec=-08^\circ31\arcmin07\arcsec.5$ (J2000).

This accuracy, however, is dominated by the error in alignment of the {\it UKIRT} images to an absolute World Coordinate System (WCS; in this case 2MASS). A more accurate result was obtained by performing simple relative astrometry between the {\it UKIRT} and the \cxo imaging, providing a position accurate to $\sim 0.3\arcsec$ (using the relative location of \sgr on the {\it UKIRT} frames). Although there is no obvious counterpart for \sgr in the {\it UKIRT} images, our deepest plausible limits come from our {\it VLT/HAWKI~} observation with a total exposure of 2640 seconds. The very small field of view (FoV) of these observations precluded performing astrometry with \cxo sources directly (none of the five X-ray sources were in the {\it HAWKI} FoV). However, we could perform relative astrometry between the {\it UKIRT} and the {\it VLT~} images with a minimal error ($<$0.05$\arcsec$), and hence place \sgr on the {\it VLT/HAWKI~} images with $\sim 0.30\arcsec$ accuracy. The resulting location of \sgr is shown in Figure \ref{fig:image_gtc_vlt}.
 
\subsubsection{IR limits}

We do not detect a source within the $1\sigma$ {\it Chandra} error circle of \sgr in the {\it VLT/HAWKI} images. Performing photometric calibration relative to 2MASS yields an upper limit for the \sgr counterpart within the {\it Chandra}/ACIS-I error circle as $K_{\rm s} > 22.4$ (3$\sigma$). We should note, however, that this is an idealized scenario since confusion between faint sources is an issue at the lowest flux levels. While there are no sources within the $1\sigma$ {\it Chandra} error radius, there are two sources at the $\sim 1.5\sigma$ level from the {\it Chandra} position of \sgrnos: a moderately bright source (marked A in Figure \ref{fig:image_gtc_vlt}) with $K_{\rm s} = 18.2 \pm 0.1$, and a fainter source (marked B) with $K_{\rm s} = 21.2 \pm 0.2$. Since neither fall within the {\it Chandra} location, and due to the crowded nature of the field, we cannot make any strong claims for their association with \sgrnos. Future observations to search for variability of these sources may provide a stronger diagnostic.

\subsection{Pulse Timing Analysis}\label{sec:timingrxte}

We used an epoch folding algorithm to determine the pulse ephemeris of \sgrnos,  as follows. We first determined the source pulse frequency using closely spaced {\it RXTE}/PCA observations on 2010 March 21 with a total exposure of 44 ks and generated a template pulse profile in the 2$-$10 keV band. We then grouped all available PCA observations into 22 discrete time segments, each separated by at least 0.3 days. For each segment, we generated the pulse profile and cross-correlated it with the template to obtain the phase drift with respect to the template. We finally fit the phase shifts with a first or higher order polynomial to determine the pulse ephemeris of the source. The method summarized here is described in detail by \citet{woo02}.

\begin{figure}[ht]
\vspace{0.2in}
\centerline{
\plotone{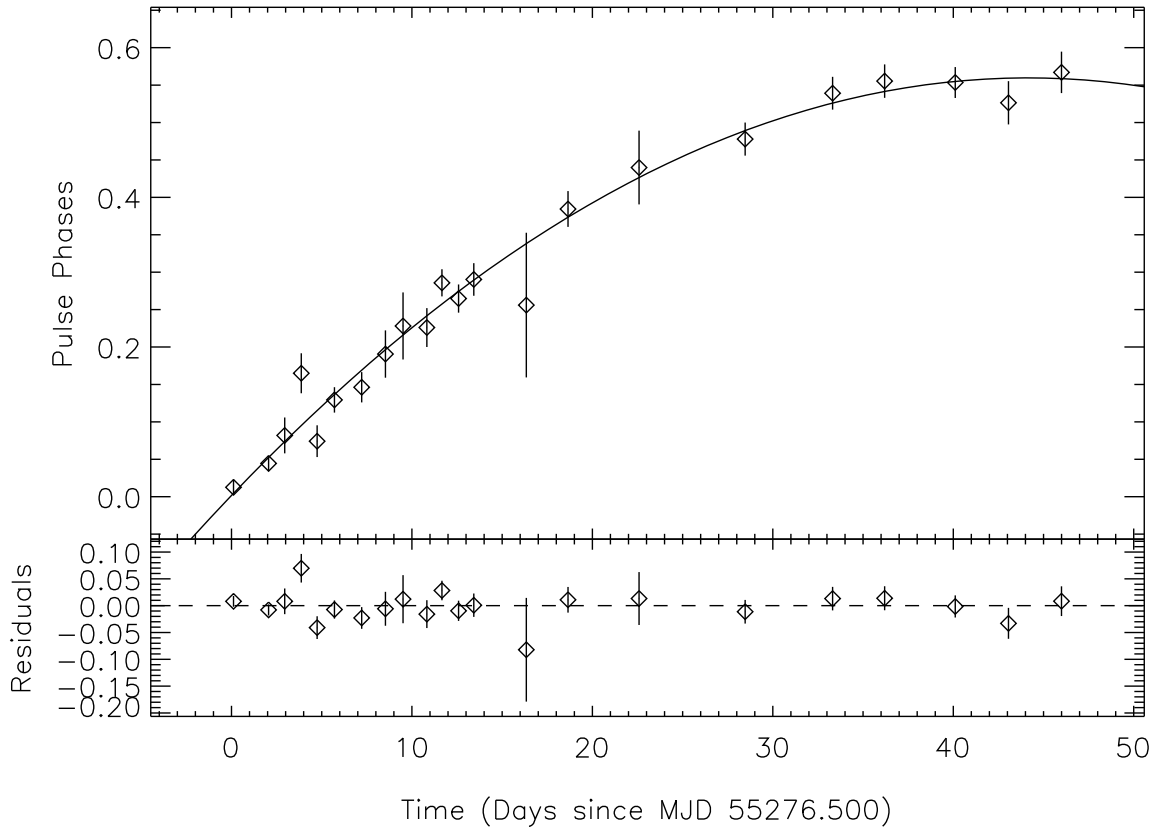}}
\caption{({\it top}) Plot of phase drifts for each segment of the PCA observations. The solid line is a second order polynomial model for the phase drift in time; ({\it bottom}) Fit residuals.}
\label{fig:phase_fit}
\end{figure}

We first attempted to model the phase drifts with respect to the template using a first order polynomial. This resulted in a poor fit ($\chi$$^2$/d.o.f. = 122.1/21) with significant deviations of phase drift values from the model. Modelling the phase drifts with a second order polynomial results in a good fit ($\chi$$^2$/d.o.f. = 19.9/20) with a spin period of 7.5654091(8) s and a period derivative of (4.39$\pm$0.43)$\times$10$^{-12}$ s/s (epoch MJD 55276.5). We present in Figure \ref{fig:phase_fit} the pulse phase drifts with respect to the phase of the template profile with a second order polynomial fit (top panel), and the residuals of the fit (lower panel).

\subsection{Pulsed Fraction and Pulsed Intensity History}

\begin{figure}[ht]
\vspace{0.2in}
\centerline{
\plottwo{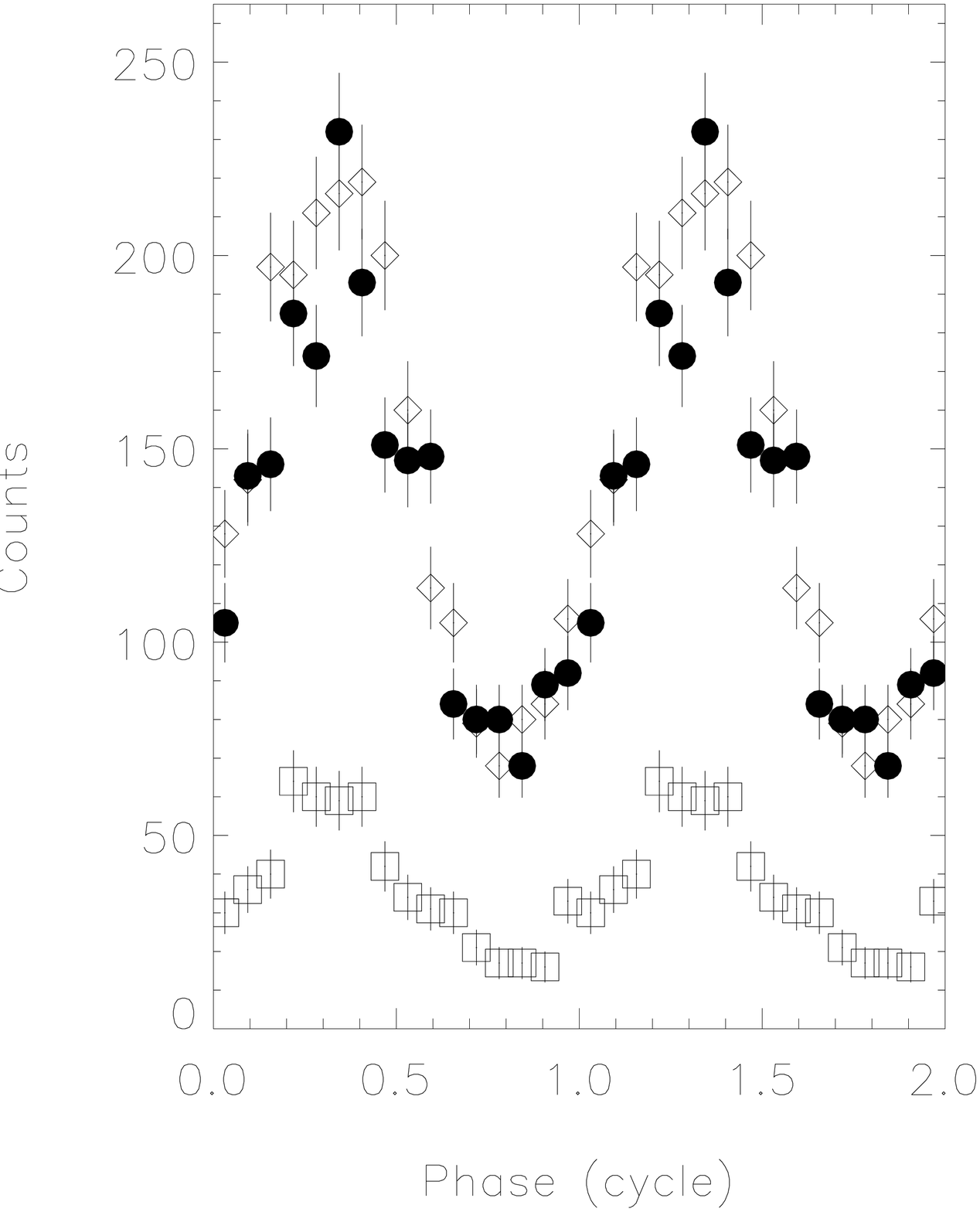}{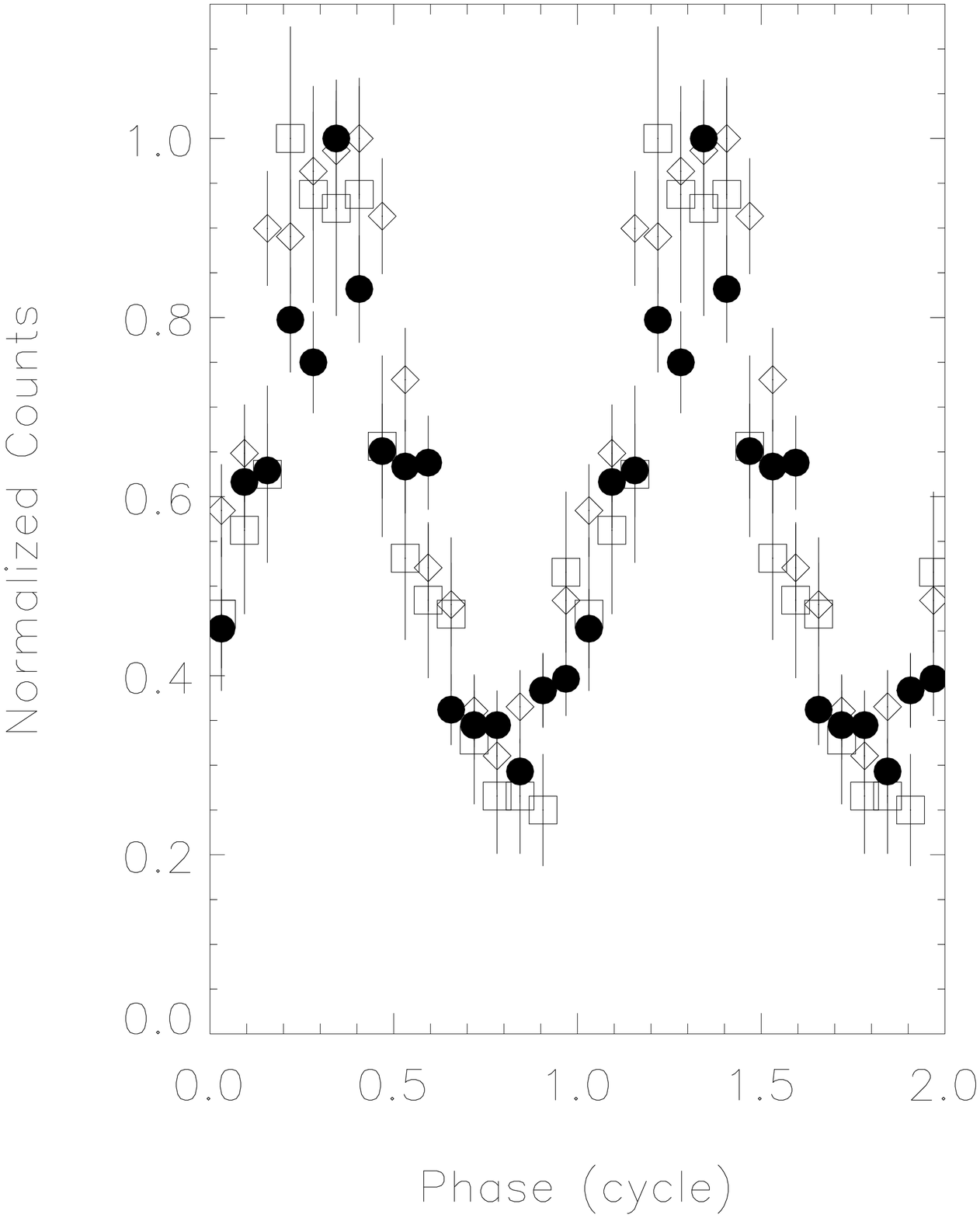}}
\vspace{0.3in}
\caption{({\it left}) Pulse profiles of \sgr in the 0.2$-$4 keV (filled circles), 4$-$6 keV (diamonds) and 6$-$10 keV (squares) energy bands; ({\it right}) normalized pulse profiles to the peak value of each energy interval.
\label{fig:pc_pps}}
\end{figure}

We investigated the energy dependence of the RMS pulsed fraction\footnotemark
\footnotetext{The RMS pulsed fraction is defined as
${\rm PF}_{\rm RMS} = \left( \frac{1}{\rm N} ( \sum_{i=1}^{\rm N}
({\rm R}_{\rm i}-{\rm R}_{\rm ave})^2 -
\Delta {\rm R}_{\rm i}^2) \right)^\frac{1}{2}$ / ${\rm R}_{\rm ave}$
, where N is the number of phase bins (N=16), ${\rm R}_{\rm i}$ is the rate in each phase bin, $\Delta {\rm R}_{\rm i}$ is the associated uncertainty in the rate, and ${\rm R}_{\rm ave}$ is the average rate of the pulse profile.} of \sgr by folding all available XRT data in PC mode in the 0.2$-$4 keV, 4$-$6 keV and 6$-$10 keV bands using the pulse ephemeris presented in \S 3.2. In Figure \ref{fig:pc_pps}, we present the pulse profiles of \sgr in the three specified energy intervals (left) and the normalized profiles obtained by dividing with the maximum value of each pulse profile (right). We note that although the intensity of pulsations varies significantly in energy, the normalized amplitude of the pulsed signal remains remarkably constant. Consequently, we find that the RMS pulsed fraction of \sgr does not vary as a function of energy with RMS pulse fractions of (34$\pm$2)\%, (36$\pm$2)\% and (39$\pm$5)\% in the above energy intervals, respectively. Therefore, combining all available data in the 0.2$-$10 keV range yields an RMS pulsed fraction of (36$\pm$2)\%,  which also remained constant over the entire course of the $\sim$30 days of XRT observations.

We also investigated the RMS pulsed intensity variations of \sgr in time and energy using all available PCA data in the 2$-$4.5 keV, 4.5$-$8.5 keV and 8.5$-$14 keV intervals. These energy bounds are selected to accommodate enough counts in each interval and determine the RMS pulsed intensities with at least 5$\sigma$ significance. We find that the pulsed count rates of the source remain fairly constant in all three energy intervals at around 0.017, 0.025 and 0.006 counts/s/PCU, respectively.

\subsection{X-ray Spectral Properties}

We performed spectral analysis using only the XRT/PC observations, to avoid the high background effects of the WT observation. The data were rebinned to 20 counts per spectral bin to improve the fitting statistics. We fit the persistent source data with single component models and found that the first 12 PC mode observations (i.e., observations 000 through 013 excluding 002 in Table \ref{sgr:tab:xrtobs}) are fully consistent with a constant source. Consequently, we summed all 12 observations into one single spectrum, which contained $\sim 5000$ counts collected over 150 ks. We then rebinned the spectrum to 30 counts per bin and limited our spectral analysis to the $2-10$ keV energy range, to account for the high absorption towards the source direction, and fit several absorbed (using {\sc phabs}) single component models within {\sc xspec}. The best fit spectral model parameters are listed in Table \ref{sgr:tab:spectrum}. Given the very high column density, the overall spectrum can be equally well fit by a steep power law ($\Gamma=3.4$) or by a blackbody model with $kT\sim1$ keV; no additional components are required. We note that the derived neutral hydrogen column densities (see Table \ref{sgr:tab:spectrum}) are larger than the estimated Galactic column density of $1.7\times 10^{22}$ cm$^{-2}$ in the direction of \sgrnos. When we fixed the column density to this Galactic value, no single or double component model was able to account for the observed spectrum, indicating that the high column density could be intrinsic to the source or to its surroundings.
It is important to also note that the origin of the high column density might be high density of molecular hydrogen in the Galactic plane, on which the source is located.  
  
\begin{table}[h]
\begin{center}
\caption{Results of the spectral fits to the first 12 PC mode {\it Swift}/XRT observations.\label{sgr:tab:spectrum} }
\begin{tabular}{ccccc}
\hline
\noalign{\smallskip}
Model & Column density      & $\Gamma$/$k\,T$& 2--10 unabs.(abs) flux             & $\chi^2_{\rm red}$   \\ 
      &($10^{22}$ cm$^{-2}$)& --/(keV)      &($10^{-12}$ erg cm$^{-2}$ s$^{-1}$) & (dof)                 \\
\noalign{\smallskip}
\hline
\noalign{\smallskip}
Power law &$16.4^{+1.1}_{-1.0}$ & $3.4^{+0.2}_{-0.2}$  & 17.3 (4.0) & 1.03 (143)\\
Black Body$^*$&$9.9^{+0.7}_{-0.7}$  &$1.11^{+0.05}_{-0.05}$&  7.4 (3.8) & 1.06 (143)\\
Bremsstr. &$13.7^{+0.8}_{-0.7}$ &$3.0^{+0.4}_{-0.3}$   & 11.6 (3.9) & 1.01 (143)\\
\noalign{\smallskip} 
\hline
\end{tabular}
\end{center}
*The equivalent black body radius for a source at 10 kpc is $R_{\rm bb}=0.7\pm0.1$ km.
\end{table}

We find evidence that the X-ray spectrum of \sgr hardens as its flux starts to decline in the last four PC mode observations (014 through 017 in Table \ref{sgr:tab:xrtobs}): an absorbed power law fit to these pointings yields indices in the interval from 2.6$-$3.2, and a black body fit yields temperatures in the interval from 1.2$-$1.3 keV (both at 90\% confidence). The corresponding 2$-$10 keV unabsorbed (absorbed) power law and black body fluxes are 12.1$\times10^{-12}$ (3.5$\times10^{-12}$) erg cm$^{-2}$ s$^{-1}$ and 5.8$\times10^{-12}$ (3.2$\times10^{-12}$) erg cm$^{-2}$ s$^{-1}$, respectively.

\section{Discussion}

\subsection{Burst emission}

At the onset of its active episode, \sgr emitted a single, relatively faint, burst that was seen with {\it Swift}/BAT and INTEGRAL \citep{kui10}; the total event fluence (15$-$350 keV) was 1.3$\times$10$^{-8}$ erg/cm$^2$. Compared to the persistent emission flux level (2$-$10 keV), the burst energetics are very low: it would take only 13 minutes for the persistent source to emit an equivalent amount of energy. It is, therefore, unlikely that the (detected) burst emission provided the energy budget for the persistent X-ray emission. 

The spectrum of the {\it Swift}/BAT burst is well described by a power law with exponential cutoff (E$_{peak}$ = 38 keV) or a black body with kT=10 keV. Its spectral and temporal characteristics are very similar to those of known magnetars, in particular, to those of SGR~J0418$+$5729, which emitted only two rather dim bursts in one day \citep{van10}, and 1E 2259$+$586, which displayed a short episode of bursting for about 10 ks in 2002 seen only in hard X-rays \citep{kas03}. In addition to the {\it Swift}/BAT event, we identified 4 weak X-ray bursts in our monitoring observations with the {\it RXTE}/PCA, which were not detected in the simultaneous {\it Fermi}/GBM data. It is interesting to note that the {\it RXTE}/PCA bursts are very similar in duration and X-ray intensity to the five short and rather weak bursts, also detected in the {\it RXTE}/PCA passband, emitted by the young, relatively strongly magnetized (B$_{dipole}$=4.9$\times$10$^{13}$ G) rotation powered pulsar PSR J1846$-$0258 \citep{gav09}.

\subsection{Persistent emission properties}

The spin period, 7.5654091 s,  and period derivative, 4.39$\times10^{-12}$ s/s, of the source derived using the {\it RXTE} data are consistent (within 1.5$\sigma$) with the values obtained by \citet{esp10}. Using these values, we estimate the dipole magnetic field strength of \sgr to be  B = 1.8 $\times$10$^{14}$ G, which is at the lower end of the B-fields of SGRs and close to the median value of those of AXPs (B$_{\rm median}$ = 2.2$\times$10$^{14}$ G). Note that the two recently discovered SGRs J$0418+5729$ and J$0501+4516$ possess the lowest SGR surface magnetic fields of $<$ 3$\times$10$^{13}$ G and 1.7$\times$10$^{14}$ G, respectively. 

\begin{figure}[h]
\vspace{0.0in}
\centerline{
\includegraphics[angle=-90,width=6.0in]{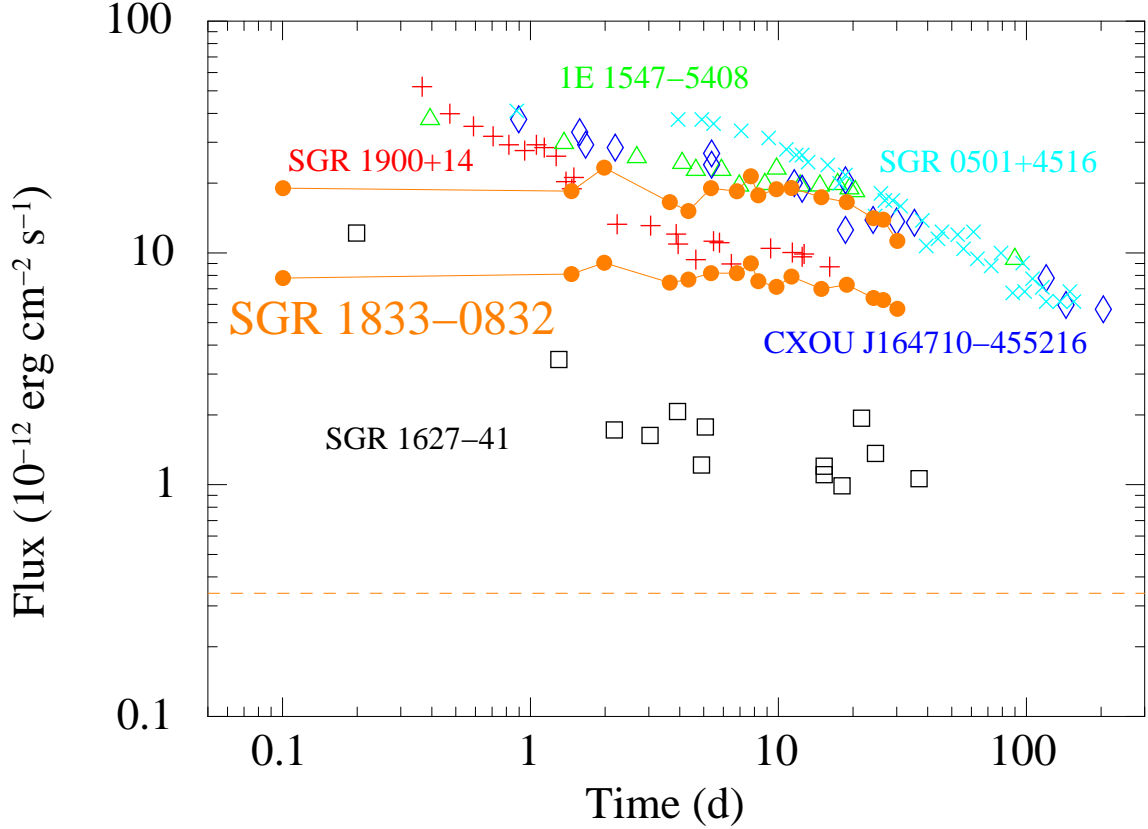}}
\caption{Unabsorbed flux of \sgr (filled circles) obtained by fitting power law (upper values) and blackbody (lower values) models to XRT data in PC mode, compared with other sources [SGR 1900+14 (plus), 1E 1547.0$-$5408 (triangles), SGR 0501+4516 (cross), CXOU J164710$-$455216 (diamonds) and SGR 1627$-$41 (squares); data obtained from \citealt{rea09}]. Time is measured from the onset of burst active episodes. The horizontal dashed line indicates the {\it Chandra} 2$\sigma$ flux upper limit of \sgr from 2009 February 13.}
\label{fig:xrtflux}
\end{figure}

Figure \ref{fig:xrtflux} exhibits the evolution of the persistent X-ray emission unabsorbed fluxes ($1-10$ keV) for five magnetars (data obtained from \citealt{rea09}) and \sgrnos, observed with {\it Swift}/XRT in the $2-10$ keV range. Each source was observed immediately after an outburst, when the persistent emission level had significantly increased from its quiescent state. We notice that all sources had similar maxima at the onset of their activation, ranging between $10^{-11}$ and $10^{-10}$ erg/cm$^2$/s. Three of the sources (SGRs~J$1627-41$, J$0501+4516$ and J$1900+14$) declined in a similar monotonic manner. However, SGR~J$1627-41$ settled at the lowest flux level of all sources ($\sim10^{-12}$ erg/cm$^2$/s), while SGR~J$1900+14$ reached a flat plateau an order of magnitude higher and SGR~J$0501+4516$ was still declining at the cessation of the XRT observations. The other two sources (1E~$1547-5408$ and CXOU~J$164710-455216$) show similar trends (but different from the other three sources): they remain flat after activation for $10-20$ days and decline {\it thereafter} with the same trend as the former three.

We have plotted the \sgr data as a band with an upper and lower bound on Figure \ref{fig:xrtflux}; the upper flux limit is derived when we fit a single power law and the lower with a blackbody fit. Because of the excessive column density towards the source, there are very few photons below 2 keV and the spectral fits are not well constrained; the real unabsorbed flux of the source is within these two bounds, and seems to remain constant over the first $\sim$20 days following the onset. 
Interestingly, we found an archival {\it Chandra}/ACIS-S observation covering the field of \sgr starting on 2009 February 13 10:50:20 UT for a total of 8 ks. The source was not detected within our \cxo error circle (\S\ref{sec:chandraobs}) and we were able to derive a $2\sigma$ upper limit (indicated by a dotted line on Figure \ref{fig:xrtflux}) for its quiescent unabsorbed flux of $3.4\times 10^{-13}$ erg/cm$^2$/s \citep{gogus10b}. This is almost 100 times fainter than the current state of the source; assuming that \sgr reaches this level again, it would take at least 2 years if it follows the same decline trend of the other five sources as indicated by the last XRT observations. The source quiescent level is close the lowest flux level of SGR 1627$-$41, the faintest magnetar thus far (Mereghetti et al. 2006, Esposito et al. 2009), and also to the flux of the faintest dim isolated neutron star RX J0420.0$-$5022 (Haberl et al. 2004) of 4.8 $\times$ 10$^{-13}$ erg/cm$^2$/s. 

Finally, pulsed radio emission has been detected from two magnetars, XTE\,J1810-197 \citep{camilo2006} and 1E\,1547.0-5408 \citep[SGR\,J1550-5418;][]{camilo2007}. The former showed radio pulsations after an X-ray outburst from the source, but the latter was seen in radio while the source was quiescent in X-rays. Searches for pulsed radio emission from other magnetars have been unsuccessful \citep[e.g.][]{burgay2006,crawford2007,hessels2008}. Our search for radio emission from \sgr with the {\it WSRT} resulted in non-detections, with upper limits of $0.2-0.6$ and $0.1-0.3$,mJy (depending on the assumed pulse widths) at 2280 and 1380\,MHz, respectively. \citet{burgay2010} reported an upper limit of 0.09 mJy at 1400 MHz on 2010 March 25 with the Parkes radio telescope. These upper limits are somewhat higher than some of the deepest radio searches for other magnetars, but more than an order of magnitude lower than the detections of XTE\,J$1810-197$ and 1E\,$1547.0-5408$. 

\subsection{Environment of \sgr}
 
A visual inspection of the region surrounding \sgr on the $K_{\rm s}-$band images of the {\it UKIRT~} GPS reveals no obvious source clustering in the vicinity of the source. However, as the field is small and crowded, low mass open clusters may be difficult to discern from bright background sources. We have performed a color-color search to uncover any cluster members; we caution, however, that it is still likely that the number of ``unrelated'' sources may obscure the detection of any faint, low surface brightness cluster that may be present.
 
There are several sources in the larger field which are plausible sites for the birth of \sgrnos. One of these is the HII region IRAS~$18310-0825$, at $RA=18^{\rm h}33^{\rm m}47^{\rm s}.9$, $Dec = -08^\circ23\arcmin52\arcsec$, which is offset by roughly $7\arcmin$ from \sgrnos, another is [LPH96] 023.162+0.023 ($RA = 18^{\rm h}33^{\rm m}23^{\rm s}.9, Dec=-08^\circ40\arcmin 32\arcsec$; 10 arcmin offset, \citealt{loc96}), while a third is [KC97c] G023.5+00.0 at $RA= 18^{\rm h}34^{\rm m}20^{\rm s}.0, Dec=-08^\circ 21\arcmin23\arcsec.7$, which is also associated with a cluster known as BDS2003 118 \citep{loc86}. Further, there are several dark clouds in the vicinity of \sgrnos, which apparently do not yet host stellar populations. They have masses of $10^2-10^3$ M$_{\odot}$, and are likely locations for a next generation of star formation within the region. Finally, there is also a known pulsar (PSR J1830-08) roughly 3'7 away from SGR J1833-0832 with an adopted distance of 5.67 kpc and a characteristic age of 147 kyr. It is possible that this source originates from the same star forming region associated with \sgrnos.
 
While the dynamics of SGRs are as yet effectively unconstrained by observations (no SGRs have yet proper motion or radial velocity measurements) we can consider constraints by assuming that they i) have a short lifetime $<10^4$ years, and ii) have a kick velocity distribution broadly comparable to that seen for young pulsars. In practice this second assumption only improves the constraints we can make if the actual kicks are {\em significantly} smaller, and hence limit the transverse distance that \sgr could have travelled since birth. A typical pulsar kick might be of order 200 km s$^{-1}$, so that in $10^4$ years the source may move $\sim 2$ pc on the sky. Hence the angular scale \sgr could cover would be $\theta \approx 400\arcsec / d_{kpc}$. For a distance of 6-12 kpc this corresponds to an angular scale of $30-70\arcsec$, although allowing for a range of kick velocities might extend this range by a factor of 3 or more. In this case it would seem that none of the candidate clusters above are likely birthplaces. However, it is also possible (although not necessarily likely) that any progenitor may have ``evaporated'' from the cluster in which it formed with a velocity dispersion of a few km s$^{-1}$. With a lifetime of a few Myr, this may result in a progenitor at a large distance from the cluster, making our constraints weaker.

\acknowledgments 

EG and YK acknowledge EU FP6 Transfer of Knowledge Project ``Astrophysics of Neutron Stars'' (MTKD-CT-2006-042722). JG and DNB acknowledge support from NASA contract NAS5-00136. AJvdH was supported by an appointment to the NASA Postdoctoral Program at the MSFC, administered by Oak Ridge Associated Universities through a contract with NASA. CK acknowledges support by NASA grant NNH07ZDA001-GLAST. Partly based on observations made with the Gran Telescopio Canarias (GTC), instaled in 
the Spanish Observatorio del Roque de los Muchachos of the Instituto de Astrofisica de Canarias, in the island of La Palma. The WSRT is operated by ASTRON (Netherlands Institute for Radio Astronomy) with support from the Netherlands foundation for Scientific Research. Partly based on observations made with ESO Telescopes at the La Silla or Paranal Observatories under programme ID 084.D-0621.

\end{document}